\documentclass[twocolumn]{revtex4}

\usepackage{amsmath}
\usepackage{graphicx}
 
\usepackage{subfigure}
\usepackage{psfrag}

\newcommand{\tr}{\mathrm{tr}\,}         
\newcommand{\del}{\partial}             
\newcommand{\half}{\frac{1}{2}}
\newcommand{\halfs}[1]{\frac{#1}{2}}
\newcommand{\expect}[1]{\left\langle #1 \right\rangle}
\newcommand{\ord}[1]{\mathcal{O}\left( #1 \right)}
\newcommand{\ordN}[1]{\ord{ N^{#1} }}
\newcommand{\oN}[1]{ \frac{1}{ N^{#1} } }
\newcommand{\Zb}[1]{\bar{Z}_{#1}}
\newcommand{\At}{\tilde{A}}
\newcommand{\Vh}{V^{\half}}

\psfrag{DD}{$D$}
\psfrag{Gb}{$E$}
\psfrag{KK}{$K$}
\psfrag{LL}{$L$}
\psfrag{Delta}{$\Delta$}
\psfrag{SS}{$\Sigma$}
\psfrag{ZZ}{$Z$}
\psfrag{Zb}{$\bar{Z}$}
\psfrag{G}{$G$}
\psfrag{GG}{$\Gamma$}
\psfrag{Th}{$\Theta$} 
\psfrag{==}{$=$} \psfrag{++}{$+$} \psfrag{cdots}{$\cdots$}
\psfrag{1N}{\scalebox{1.8}{$\frac{1}{N}$}}
\psfrag{xx}{$\times$}
\psfrag{ii}{$i$} \psfrag{ii++}{$i+1$} \psfrag{ii--}{$i-1$}
\psfrag{jj}{$j$}
\psfrag{kk}{$k$}
\psfrag{ll}{$l$}
\psfrag{mm}{$m$}
\psfrag{nn}{$n$}
\psfrag{oo}{$o$}
\psfrag{pp}{$p$}
\psfrag{qq}{$q$}
\psfrag{11}{1}
\psfrag{1}{1} 
\psfrag{LL++}{$L+1$} \psfrag{LL--}{$L-1$}
\psfrag{JJ}{$J$}
\psfrag{A}{$A$}
\psfrag{C}{$C$}
\psfrag{U}{$U$}

\begin{document}

\title{An Algorithm for RNA Pseudoknots}

\author{M. Pillsbury}
\affiliation{Department of Physics, University of California, Santa 
Barbara, CA 93106, USA}  
\author{J. A. Taylor}
\affiliation{Department of Physics, University of California, Santa
Barbara, CA 93106, USA} 
\author{H. Orland}
\affiliation{Service de Physique Th\'eorique, CEA Saclay, 91191
Gif-sur-Yvette Cedex, France} 
\author{A. Zee}
\affiliation{Institute of Theoretical Physics, University of
California, Santa Barbara, CA 93106, USA}
\affiliation{Department of Physics, University of California, Santa
Barbara, CA 93106, USA} 

\begin{abstract}
  We further develop the large $ N $ formalism presented by some of us
  in earlier works in order to recursively calculate the partition
  function of a singly pseudoknotted RNA. We demonstrate that this
  calculation takes time proportional to the sixth power of the length
  of the RNA. The algorithm itself is presented in a self-contained
  form for the convenience of readers interested in implementing it.
\end{abstract}

\maketitle 

\section{Introduction}
\label{sec:intro}

\noindent
An RNA molecule is a heteropolymer, consisting of four types of
nucleotides---uracil ($U$), adenine ($A$), guanine ($G$) and cytosine
($C$)---which are linked into a chain by phosphate bonds. The
nucleotides then form saturating hydrogen bonds with one another, with
$ A $ bonding to $ U$ and $ C $ bonding to $ G$. These bonds cause the
RNA to fold up into a stable, three dimensional structure. The
structure is further stabilized by the stacking interactions, where
runs of adjacent bonds twist into the familiar Watson-Crick
helices. 

The folding process is believed to be roughly
hierarchical\cite{higgs,tinoco}, in that the primary (one-dimensional)
structure, defined as the sequence of nucleotides, determines the
secondary (two-dimensional) structure. If one visualizes the RNA as a
circle, tying the first nucleotide to the last, and then represents
hydrogen bonds as arcs within the circle, the secondary structure is
the structure that can be drawn without bonds crossing one
another\cite{nj,oz}. The secondary structure is largely unaffected by
the formation of the (three-dimensional) tertiary structure, which
includes, among other things, pseudoknots.  Pseudoknots are precisely
those structures which \emph{do} contain bonds that would cross each
other inside the disk. Algorithms that allow for the prediction of
certain pseudoknots exist. For instance, \cite{rivas} allows for any
number of bonds on the outside of the disk, as long as they do not
cross. Our algorithm allows for a different set of topologies,
enumerated in \cite{poz}.

We only consider a basic model of RNA, in which we only account for
the energies associated with bond formation and neglect stacking
interactions, loop entropies, and steric constraints.  Modern RNA
folding algorithms tend to be more complex, especially towards
secondary structure\cite{zuker,jaeger}, but we opt for a simpler
approach for the sake of clarity in this paper. As a result, the
implementation we outline here is merely an illustration; it is not
useful for predicting the structure of folded RNA. 

Let $-U_{ij}$ be the change in energy of forming a hydrogen bond
between the $i$th and $j$th bases, and let $V_{ij}= \exp (U_{ij}/ T)$
be the Boltzmann factor at temperature $T$.  Then one can write the
partition function as a sum of terms of the form $ V_{i_1i_2}
V_{i_3i_4} \dots V_{i_{n-1} i_n} $.  The $ i_k $s in the above sum
must be unique because a base is saturated by a single hydrogen bond.
Our partition function is then
\begin{multline}
  \label{eq:pt-fn-v}
  Z_{L1} = 1 + \sum_{i<j} V_{ij} 
    + \!\!\!\! \sum_{i < j < k < l }  \!\!\!\! V_{ij} V_{kl} \\
    + \dots + \oN{2}\!\! \!\! \sum_{i < j < k < l}  \!\!\!\!
    V_{ik} V_{jl} + \dots
\end{multline}
The factor of $ N $ is a parameter which tracks the topology of the
bond structure. Since there is usually a small number of pseudoknots,
and they can be controlled by the concentration of $ \mathrm{Mg}^{++}
$ ions in solution, we have introduced the parameter $ N $, and
penalize terms involving pseudoknots with factors of $ 1/N^2 $. The
secondary structure is $ \ord{1} $, while a restricted set of
singly-pseudoknotted structures are $\ord{N^{-2}}$\cite{oz}.

There are many ($\sim L!$) possible terms in the above sum. We propose
an algorithm which allows us to calculate this partition function to $
\ord{N^{-2}} $ in $ \ord{L^6} $ time. We also generalize this
algorithm to calculate the partition function of an RNA that contains
an arbitrary number of these $ \ord{N^{-2}} $ pseudoknots. We can then
use backtracking to determine what bonds contribute to the dominant
term in the partition function, and thus the folded structure at low
temperatures. We can also determine the probability that a given bond
is present at a finite temperature.

In order to express sums like (\ref{eq:pt-fn-v}) in a convenient
fashion, we introduced two objects, which we call ``propagators''. The
first of these is the two-indexed $G_{ij}$, which is the partition
function of the secondary structure of a chain of nucleotides
identical to that which lies between the $ j $th base and the $ i $th
base. We set $ G_{i,i+1} = 1$ and $ G_{i,i+l} =0 $ for $ l > 1 $,
which forbids ``backwards propagation'', insuring $ G_{ij} $ always
runs in the $3'$ to $5'$ direction. The other propagator is the
four-indexed $ \Delta_{ij;kl} $, which is non-zero if $ i \ge k > l
\ge j $. In that case, $ \Delta_{ij;kl} $ is the partition function of
the structure of two spatially separated, anti-parallel, chains of
nucleotides, one between $ k $ and $ i $, and the other between $ j $
and $ l $, .

\begin{figure}[hbt!]
  \centering
  \subfigure[Diagram for $ G_{ij} $.]{
    \label{fig:el-prop:di-g}
    \scalebox{.8}{\includegraphics{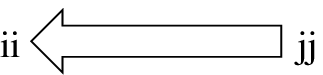}}}\
  \hspace{0.2in}
  \subfigure[Typical structure  $ G_{ij} $.]{
    \label{fig:el-prop:ty-g}
    \scalebox{.8}{\includegraphics{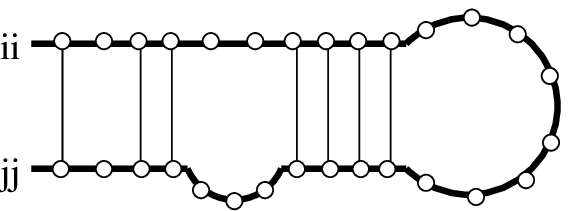}}}
  \hspace{0.2in}
  \subfigure[Diagram for $ D_{ij;kl} $.]{
    \label{fig:el-prop:di-d}
    \scalebox{.8}{\includegraphics{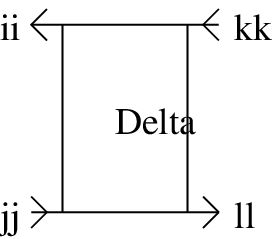}}}
  \hspace{0.2in}
  \subfigure[Typical structure $  D_{ij;kl} $.]{
    \label{fig:el-prop:ty-d} 
    \scalebox{.8}{\includegraphics{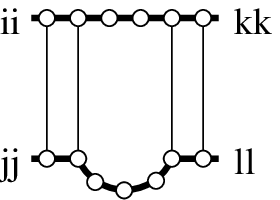}}}
  \caption{$ G $ and $ \Delta $.}
  \label{fig:el-prop}
\end{figure}

These definitions allow us to depict topological classes of structures
with Feynman diagrams. In these diagrams, $ G_{ij} $ is an arrow
pointing from $ j $ to $ i $, and $ \Delta_{ij;kl} $ is a box with
solid sides with corners at $i$, $j$, $k$, and $l$. Individual $
V_{ij} $ bonds are dashed lines from $ i $ to $j $. These diagrams are
typically in one-to-one correspondence with sums, as in \cite{poz}.
The diagrams, as well as typical associated structures, are shown in
fig.~\ref{fig:el-prop}.  In figs.~\ref{fig:el-prop:ty-g} and
\ref{fig:el-prop:ty-d}, the heavy lines represent the RNA backbone,
and the empty circles represent nucleotides.

We begin by summarizing the results relevant to a computer
implementation of the algorithm are in section~\ref{sec:algo}. We
present two variations on the algorithm.  One allows for the
calculation of the partition function with a single pseudoknot. The
other allows for any number of pseudoknots. The remainder of the paper
presents derivations of these results for the interested reader.

The index structure makes the topology of RNA folding similar to that
of large $N$ quantum chromodynamics, and these diagrams can be
expressed using the t'Hooft double line formalism\cite{thooft,
  coleman}. We review the derivations of the recursive equations which
define $ \Delta $ and $ G $, as well some other techniques and
notations from \cite{oz} and \cite{poz}, in section~\ref{sec:matrix}.

Calculating the pseudoknot partition function is considerably more
involved than backtracking the secondary structure partition
function. Instead of a single, simple sum, many topologies must be
summed over, and many of those involve summations over several
indices. In order to do these sums efficiently, we define new
propagators, similar to $ G $ and $ \Delta $, in section
\ref{sec:props}. 

The procedure for calculating the partition function remains a
recursive one, though. One finds the partition function $ Z_{L+1,1} $
of an $ L+1 $ nucleotide RNA in terms of the structures of the $ L $
nucleotide RNA. This is simple enough for the secondary structure,
because it is defined to exclude crossings. Thus, one considers only
structures where a single bond arcs over complete secondary
structures. This reasoning will lead one to the defining relation for
$ G $ described in section~\ref{sec:matrix}, and, indeed, it is how
the equation was first derived\cite{nj}.

\begin{figure}[hbtp]
  \centering
  \scalebox{.6}{
    \includegraphics{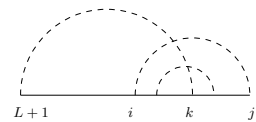}}
  \caption{The bond from $ L + 1 $ to $ k $ is allowed when
    considering pseudoknots.} 
  \label{fig:ps-vert}
\end{figure}

However, crossings are the defining feature of pseudoknots, so our
recursion relation will need to handle them. This requires the
introduction of the ``vertex operators'' mentioned in section VI of
\cite{oz}. These vertex operators are defined to be equal the
partition functions of the secondary and pseudoknot structures on runs
of nucleotides, with the additional requirement that a given
nucleotide is unsaturated, so an incoming hydrogen bond can attach
there. For example, the structure between the bases marked $ i $ and $
j $ in fig. \ref{fig:ps-vert} can be associated with a term in a
vertex operator that allows an incoming bond from the $ L+1 $st base
to connect at $ k $.

Once these vertex operators are put in a form similar to that of the
pseudoknots in \cite{poz}, which can be summed explicitly, it is
possible to use them to write a recursion relation that can be used to
generate a partition function that includes pseudoknots. Fortunately,
the techniques needed to evaluate these vertex operators are very
similar to those used to evaluate the pseudoknots. These calculations
are described in section \ref{sec:verts}.

\section{Algorithm}
\label{sec:algo}

\noindent
Using the methods and results of the previous section, one can
calculate the partition function of a singly pseudoknotted RNA with $
L $ nucleotides as follows:

\begin{enumerate}
\item Generate the matrix $ V_{ij} $ from the RNA sequence. 
\item Solve the Hartree equation
  \begin{displaymath}
  G_{ij} = G_{i-1,j} + \sum_{k=j}^{i-1} V_{ik} G_{i-1,k+1} G_{k-1,j}  
    + \delta_{i,j-1}  
  \end{displaymath}
for $ G $.
\item Use the result for $ G $ and the Bethe-Salpeter equation
  \begin{multline*}
    D_{ij;kl} = V_{ij} \bigg[ \delta_{ik} \delta_{lj}\\
    + \sum_{m=k}^{i-1} \sum_{n=j+1}^{l} \!\! 
    G_{i-1,m+1} D_{mn;kl} G_{n-1,j+1} \bigg]
  \end{multline*}
to find $ D $.
\item Use $ D $ and $ G $ to determine $ E $, $ K $, and $ L $. 
  \begin{align*}
      E_{ij;kl} & = \sum_{m=j}^l \sum_{n=k}^i D_{ij;nm} G_{n-1,k}
        G_{l,m+1} \\  
      K_{ij;kl} & = \sum_{m=j}^l \sum_{n=k}^i D_{nj;km} G_{l,m+1}
        G_{i,n+1} \\   
      L_{ij;kl} & = \sum_{m=j}^l \sum_{n=k}^i D_{im;nl} G_{n-1,k}
        G_{m-1,j} 
  \end{align*}
\item Calculate the vertex $ \Gamma_{ij}^J $, defined by
  \begin{displaymath}
    \Gamma_{ij}^J = E_{ij;J+1,J-1} 
  \end{displaymath}
  Then calculate the $ \Theta_{ij}^J $, defined by all the diagrams in
  fig.~\ref{fig:fin}. The explicit sums associated with the terms
  in $ \Theta^J_{ij} $ are contained in appendix~\ref{sec:sums}.
\item Use the recursion relation 
  \begin{multline}
  \nonumber
    Z_{L1} = Z_{L-1,1} \\   
         + \sum_{J=1}^{L-1} V_{LJ}      
         \left(Z_{L-1,J+1} G_{J-1,1} + G_{L-1,J+1} Z_{L-1,1}\right) 
         \\   
         + \frac{1}{N} \sum_{J=1}^{L-1} \sum_{mn}  
           V_{LJ} G_{L-1,m+1} \Gamma_{mn}^J G_{n-1,1} 
         \\ 
         + \sum_{J=1}^{L-1} \sum_{mn} 
           V_{LJ} G_{L-1,m+1} \Theta_{mn}^J G_{n-1,1}
  \end{multline}
to find the partition function $ Z_{L1} $.
\end{enumerate} 

Alternatively, one can find the partition function for an RNA with an
arbitrary number of pseudoknots by replacing the recursion relation in
step 5 with (\ref{eq:arb-pk}). 
\begin{eqnarray*}
  Z_{L1} & = & Z_{L-1,1} \\    
         & + & \sum_{J=1}^{L-1} V_{LJ}     
         Z_{L-1,J+1} Z_{J-1,1} \\   
         & + & \frac{1}{N} \sum_{J=1}^{L-1} \sum_{mn}   
           V_{LJ} G_{L-1,m+1} \Gamma_{mn}^J G_{n-1,1} \\ 
         & + & \sum_{J=1}^{L-1} \sum_{mn}   
           V_{LJ} G_{L-1,m+1} \Theta_{mn}^J G_{n-1,1}  
\end{eqnarray*}

\begin{figure}[htbp]
  \centering
  \scalebox{1}{
    \includegraphics{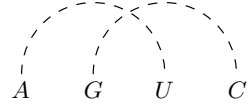}}
  \caption{Sample knot.}
  \label{fig:samp}
\end{figure}

We invite the interested reader to examine the derivation of the
algorithm in the following sections.  
The algorithm can pick out knots. An example is shown in
fig.~\ref{fig:samp}, where the (one-pseudoknot) algorithm was applied
to the sequence $ AGUC $. The arcs represent hydrogen bonds.

\section{Matrix Formalism and Steepest Descent}
\label{sec:matrix}
\noindent
In \cite{oz}, two of the authors developed a formalism involving
integrating over sets of matrices in order to exploit known results
from quantum field theory.  In this section, we review the techniques
and notations associated with this formalism that we use in this
article.

The analysis began with a demonstration that (\ref{eq:pt-fn-v}) is
equivalent to
\begin{equation} 
  \label{eq:zder}
  Z_{L1} = \frac{1}{N} \left.\frac{\del}{\del h}\right|_{h=0} 
      \int dA\,
      e^{-\frac{N1}{2}\tr A^2} \left( \det M(A) \right)^N \\
\end{equation}
\noindent 
The integral is taken over the space of all $ (L+1) \times (L + 1) $
hermitian matrices $A$. $M(A)$ is a matrix function of $A$
\begin{equation}
  M_{ij}(A) = \delta_{ij}-\delta_{i,j+1} + h \delta_{i1}   
      \delta_{L+1,j} + i\sqrt{V_{i-1,j}}\,A_{i-1,j}   
\label{eq:def-mfun}
\end{equation}
\noindent 
In matrix form, this reads
\begin{equation}
  \label{eq:def-mfun-big}
  M(A) = \left(
    \begin{array}{cccccc}
    1 & 0 & 0 & \ldots & 0 & h\\ 
    -1 & 1+a_{12} & a_{13} & \ldots & a_{1L} & 0 \\
    a_{12}^* & -1 & 1 + a_{23} & \ldots & a_{2L} & 0 \\
    \vdots & \vdots & \vdots &  \ddots  & \vdots & \vdots \\
    a_{1L}^* & a_{2L}^* & \ldots & a_{L-1,L}^* & -1 & 1 
  \end{array} 
  \right)
\end{equation}
\noindent 
Here, we have abbreviated the notation so that
\begin{eqnarray}
  \nonumber 
  a_{ij} & = & i V^\half_{ij}\,A_{ij}\qquad \mathrm{for}\; i<j \\ 
  \label{eq:aij}
  a_{ij}^* & = & i V^\half_{ij}\,A_{ij}\qquad \mathrm{for}\; j>i  
\end{eqnarray}
\noindent
Note that we use $ V^\half_{ij} $ to refer to the square root of the $
(i,j)$th element of $ V $.

We use the well-known identity $ \det M = e^{\tr\log M} $ and evaluate
the derivative in (\ref{eq:zder}) to get
\begin{equation}
  \label{eq:zexp}
  Z_{L1} = \int dA\, e^{-\halfs{N} \tr A^2 + N\tr\log M(A)}
  M^{-1} (A)_{L+1,1} 
\end{equation}
\noindent
We can define an expectation value for the function $ O(A) $ relative
to an action $ S(A) = \half \tr A^2 - \tr\log M(A) $ as
\begin{equation}
  \label{eq:def-expect}
  \expect{O} = \int dA e^{-N S(A)} O(A) 
\end{equation}
\noindent
This allows us to write the partition function in a particularly
compact form.
\begin{equation}
  \label{eq:def-z}
  Z_{L1} = \expect{M^{-1}_{L+1,1}}
\end{equation}

\begin{figure}[hbtp]
  \centering
  \subfigure[]{
    \scalebox{.56}{\includegraphics{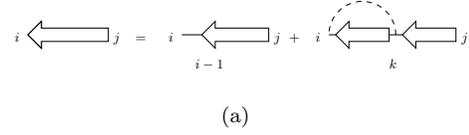}}} 
  \caption{Hartree equation for $ G_{ij} $}
  \label{fig:eq-prop:g}
\end{figure}

We can use the steepest descent approach outlined in section V of
\cite{oz} in order to calculate the quantities in brackets to
$\ordN{-2}$. We expand (\ref{eq:def-z}) in terms of a matrix
fluctuation $ x $, which is defined by
\begin{equation}
  \label{eq:def-fluc}
  A_{ij} = \tilde{A}_{ij} + \Vh_{ij} y_{ij}/N^{\half}
    = \tilde{A}_{ij} + x_{ij}/N^{\half} 
\end{equation}
\noindent
Here, $ \tilde{A}_{ij} $ is stationary point of the action $ S(A) $.
We define   
\begin{equation}
  \label{eq:g-by-m}
   G_{ij} = M^{-1}(\tilde{A})_{i+1,j}
\end{equation}
\noindent
From this, it can be shown that the propagator $ G_{ij} $ satisfies a
Hartree recursion relation.
\begin{equation}
  \label{eq:hf-g}
  G_{ij} = G_{i-1,j} + \sum_{k=j}^{i-1} V_{ik} G_{i-1,k+1} G_{k-1,j}  
    + \delta_{i,j-1}  
\end{equation}
\noindent
This can be solved after we impose boundary conditions. This equation
is familiar from the RNA folding literature, and was first used in
\cite{nj} to determine RNA secondary structure. This is to be
expected, as $ G_{ij} $ is the secondary structure partition function
for a run of nucleotides between $ j $ and $ i $.
Fig.~\ref{fig:eq-prop:g} shows the defining relation for the Hartree
propagator.

Next, we find corrections to $ G_{ij}$ by expanding $
S(\At+x/\sqrt{N}) $ as a power series in $ x $. The first term in the
action is just a quadratic,
\begin{displaymath}
  \half \tr \left(\At + \frac{x}{N^\half} \right)^2 =  
    \half \tr \At^2 + \oN{\half} \tr \At x + \frac{1}{2 N} \tr x^2 
\end{displaymath}
\noindent
Using the definitions in (\ref{eq:def-mfun}) and (\ref{eq:g-by-m}), we
can write  
\begin{widetext}
\begin{align*}
  M_{ij}\left(\At+\frac{x}{N^\half} \right) & =  M_{ij}(\At) 
      + \frac{i}{\sqrt{N}} \Vh_{i-1,j} x_{i-1,j} \\  
    & = \sum_{kl} M_{ik}(\At) \left(\delta_{kj} + \frac{i}{\sqrt{N}}
      M^{-1}_{kl}(\At) \Vh_{l-1,j} x_{l-1,j} \right) \\  
  \tr \log M \left(\At+\frac{x}{N^\half} \right) & = 
    \tr \log M(\At) + \tr \log \left( I +  \frac{i}{\sqrt{N}}
      \sum_{l} G_{i-1,k} \Vh_{k-1,j} x_{k-1,j} \right) \\
\end{align*}
\end{widetext}
\noindent
If we define $ R_{ij} = \sum_k G_{i-1,k} \Vh_{k-1,j} x_{k-1,j} $, we
can expand
\begin{displaymath}
  \tr \log \left(I +  i \frac{R}{N^\half} \right) =   
    \tr \left(i \frac{R}{N^\half} + \frac{R^2}{N} + \dots \right)  
\end{displaymath}
\noindent
Terms linear in $ x $ vanish because $ \At $ is the stationary point
of $ S(A) $, so our expectation value becomes
\begin{multline}
  \expect{O} = C \int dx\, \exp \bigg[
   -\half \left( \tr x^2 - \tr R^2 \right) \\
    - \sum_{p=3}^\infty \frac{(-i)^p}{p N^{\halfs{p}-1}} \tr R^p 
    \bigg] O    
  \label{eq:def-exp-new}
\end{multline}
\noindent
with the normalization $ C $ absorbing the Jacobian from the change of
variables and various other constants (like $ e^{-\half \tr \At^2} $). 

We reintroduce the shorthand of \cite{oz} and \cite{poz} for powers
and traces of powers of $ R $ in order to facilitate discussion of
these expectation values.
\begin{eqnarray}
  \label{eq:def-bp}
  B_p & = & R^p \\
  \label{eq:def-tp} 
  T_p & = & \tr B_p 
\end{eqnarray}
\noindent
Now we can write 
\begin{equation}
  \label{eq:m-inv}
  M^{-1}_{ij}\left(\At+ \frac{x}{N^\half}\right) =  
    \sum_k \left[\sum_{p=0}^\infty \frac{(-i)^p B_p} 
    {N^{\frac{p}{2}}}\right]_{ik} G_{k-1,j}  
\end{equation}
\noindent
using the power series for $ (1+x)^{-1} $.

We call the free (i.e., quadratic) part of the action
\begin{equation}
  \label{eq:def-s0}
  S_0(x) = \half \tr x^2 - \half \tr R^2 
\end{equation}
\noindent 
and use it to define the free expectation value,
\begin{equation}
  \label{eq:def-free-exp}
  \expect{O}_0 = \int dx e^{S_0(x)} O 
\end{equation}
\noindent
The free action can be written in terms of a kernel $ \Delta^{-1} $,
defined by
\begin{eqnarray*}
  S_0(x) & = & \sum_{i,j,k,l} x_{ij} \Delta^{-1}_{ij;kl} x_{kl} \\
  \Delta^{-1}_{ij;kl} & = & \delta_{il} \delta_{jk} - G_{l-1,i+1} 
    G_{j-1,k+1} V^\half_{ij} V^\half_{kl}   
\end{eqnarray*}
\noindent
Multiplying the four-indexed $ \Delta^{-1} $ by its inverse gives the
Bethe-Salpeter relation
\begin{multline}
  \Delta_{ij;kl} = \delta_{ik} \delta_{lj}  \\
    + \Vh_{ij} \sum_{m=k}^{i-1} \sum_{n=j+1}^{l} 
    G_{i-1,m+1} \Vh_{mn} \Delta_{mn;kl} G_{n-1,j+1}  
\end{multline}

\begin{figure}[hbtp]
  \scalebox{.56}{\includegraphics{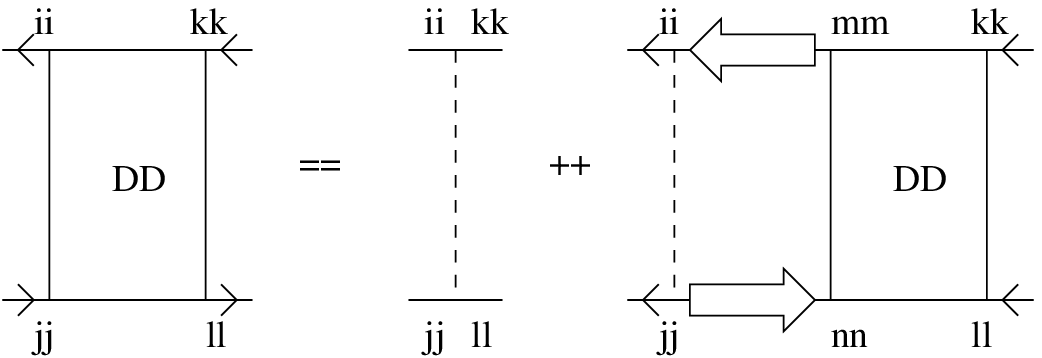}}
  \label{fig:eq-prop:d}
  \caption{Bethe-Salpeter equation for $ D_{ij;kl} $}
\end{figure}

It is convenient to multiply this equation by $ \Vh_{ij} $ and $
\Vh_{kl} $ and define a slightly different propagator
\begin{displaymath}
  D_{ij;kl} = \Vh_{ij} \Delta_{ij;kl} \Vh_{kl} 
\end{displaymath}
$ D $ obeys a new Bethe-Salpeter relation.
\begin{multline}
  \label{eq:bs-d}
  D_{ij;kl} = V_{ij} \bigg[ \delta_{ik} \delta_{lj} \\
    + \sum_{m=k}^{i-1} \sum_{n=j+1}^{l} \!\!
    G_{i-1,m+1} D_{mn;kl} G_{n-1,j+1} \bigg]
\end{multline}
The diagrammatic form of this equation is shown in
fig.~\ref{fig:eq-prop:d}.  We can use the $ D $ propagator to evaluate
expectation values by Wick contracting $x$s, or we can do a
co\"ordinate transformation in our integral $ x_{ij} = y_{ij}/\Vh_{ij}
$ and use $ D $ to contract $y$s, since the Jacobian associated with
this change of variables is just a constant which can be absorbed into
the normalization. 

Solving (\ref{eq:bs-d}) initially appears to be a task that will take $ \ord{L^6} $ time. However, the diagrammatic form of the equation indicates that $ D_{ij;kl} $ can be written as 
\begin{equation} 
\label{eq:new-g-def}
D_{i+1,j-1;k-1,l+1} = V_{i+1,j-1} V_{k-1,l+1} G_{ij;kl} 
\end{equation}
where the four-index $ G$ propagator $ G_{ij;kl} $ is calculated in the same manner as $ G_{ij} $, except the matrix $ V $ is modified so that $ V_{mn} = 0 $ if either $ l > m > k $ or  $ l > n > k $. Calculating $ G_{ij;kl} $ for a fixed $ k $ and $l $ is then a $ \ord{L^3} $ task which much be performed $ \ord{L^2} $ times, meaning that $ D $ can be calculated in $ \ord{L^5} $ time. 

Because $ G_{ij} $ represents the partition function of the possible
secondary structures on the nucleotides between $ i $ and $ j $
inclusive, we should never see anything involving $ G $ that looks
like a conventional matrix product, (e.g, $ \sum_j G_{ij} D_{jk;mn}
$), because that would include unphysical structures which have two
hydrogen bonds ending at $ j $. However, we see many terms of the form
$ \sum_j G_{ij} D_{j-1,k;mn} $, and so it is convenient, for an
arbitrary matrix $ Q $, to define a non-standard matrix product for
sums involving $ G $,
\begin{eqnarray*}
  (Q G)_{ij} & \equiv & \sum_k Q_{ik} G_{k-1,j} \\  
  (G Q)_{ij} & \equiv & \sum_k G_{i-1,k} Q_{kj} 
\end{eqnarray*}

Using these definitions we can write a more explicit form of $ Z $
\begin{multline}
  \label{eq:z-final}
  Z_{L1} = \bigg\langle \!\!\exp \bigg[  
      -\sum_{p=3}^{\infty} \frac{(-i)^p T_p} 
      {p N^{\halfs{p}-1}} \bigg]\\
      \times
    \bigg[\bigg(\sum_{q=0}^\infty \frac{(-i)^q B_q}{N^\halfs{q}}
      \bigg) G \bigg]_{L+1,1}  \bigg\rangle_0 
\end{multline}
where we have used the series expansion for $ M^{-1} ( \At+x/N^{\half} )
$. This expression can then be expanded to $ \ordN{-2} $, and the
various terms can be evaluated using Wick's theorem. 

\section{Propagators}
\label{sec:props}
\noindent
The steepest descent expansion of (\ref{eq:z-final}) was completed in
\cite{poz}, and lead to a number of sums, each of which corresponds to
a possible pseudoknot topology. While these sums need not be evaluated
in order to calculate the partition function, many similar sums must
be. As discussed in section \ref{sec:intro}, each of these
contributions will be expressed as sums involving $ D $ and $ G $
propagators, with indices running over appropriate ranges.  In order
to perform these summations efficiently, and describe them compactly,
we must introduce new propagators. These are indexed objects similar
to the already defined $ D $ and $ G $, and they also fall into two
classes, those which have two indices, and those which have four.

The first new two-indexed propagator is $ Z_{ij} $ itself, and the
second, $\Zb{mn}$, is the pseudoknot contribution---that is, the
$\ordN{-2}$ part of $Z_{mn}$.  The different two-indexed propagators
are related by the following equations,
\begin{eqnarray}
  \label{eq:def-zb}
  Z_{ij} = G_{ij} + \frac{1}{N^2} \Zb{ij}
\end{eqnarray}
\begin{figure}[htbp]
  \centering 
  \scalebox{0.8}{ 
    \includegraphics{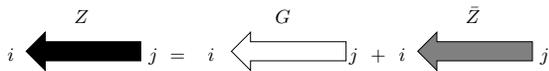}} 
  \caption{Diagrams defining $ Z $ and $ \Zb{} $.} 
  \label{fig:def-z-zb-sig}
\end{figure}

Now, one can see from the Bethe-Salpeter equation for $ D_{ij;kl} $
that it always has bonds both between the $i$th and $j$th bases and
between $k$th and $l$th bases. It is often convenient (and efficient)
to make different assumptions about which bonds are implicitly defined
when implementing a dynamic programming algorithm on a computer. To
this end, we define new four-indexed propagators, the first of which
is,
\begin{equation}
  \label{eq:def-e}
  E_{ij;kl} = \sum_{p=k}^i \sum_{q=j}^l D_{ij;pq} G_{p-1,k} G_{l,q+1} 
\end{equation}

We draw $ E_{ij;kl} $ in diagrams exactly like $ D_{ij;kl} $, except
that we draw it with a dotted edge between $ k $ and $ l $ to
represent the fact that it doesn't contain an explicit hydrogen bond
between those two nucleotides.  The physical picture of the propagator
$ D_{ij;kl} $ presented in \cite{poz}---i.e., two spatially separated
runs of nucleotides (one from $ j $ to $ l $, the other from $ k $ to
$ i $) that form nucleotide bonds with each other---is equally
applicable to $ E $. However, instead of the two strands being bound
to each other at both ends, here the inner ends of each strand may be
loose.

\begin{figure}[hbtp]
  \centering
  \scalebox{0.56}{
    \includegraphics{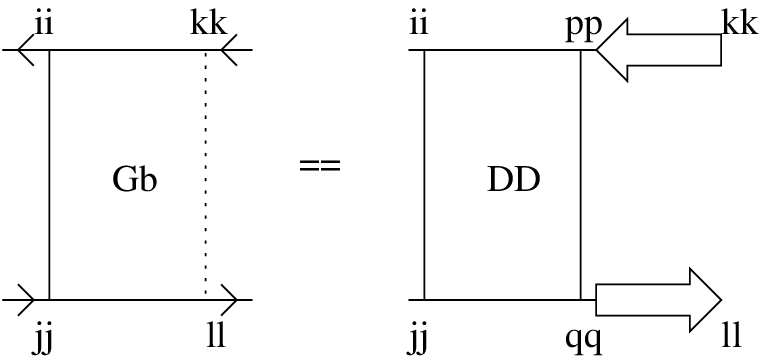}}
  \caption{Defining relation for $ E_{ij;kl} $.}
  \label{fig:def-e}
\end{figure}

It's also useful to define two other four-indexed propagators, $ K $
and $ L $. Both $ K_{ij;kl} $ and $ L_{ij;kl} $ have explicit bonds
connecting $ i $ and $ j $. However, $ K_{ij;kl} $ is defined to have
a bond connecting to $ k $, but not necessarily to $ l $. Likewise, $
L_{ij;kl} $ will always have a bond connecting at $ l $, but may not
have a bond connecting to $ k $.
\begin{eqnarray}
  \label{eq:def-k}
  K_{ij;kl} = \sum_{m=j}^l \sum_{n=k}^i D_{nj;km} G_{l,m+1} G_{i,n+1} \\ 
  \label{eq:def-l}
  L_{ij;kl} = \sum_{m=j}^l \sum_{n=k}^i D_{im;nl} G_{n-1,k} G_{m-1,j}
\end{eqnarray}
\noindent
Diagrams for $ K $ and $ L $ have a dotted edge drawn to the
nucleotide which doesn't have an explicit bond connecting to it, and
solid edges drawn to the nucleotides that do attach to bonds. 
\begin{figure}[hbtp]
  \centering
  \subfigure[$ K $]{
    \scalebox{.7}{\includegraphics{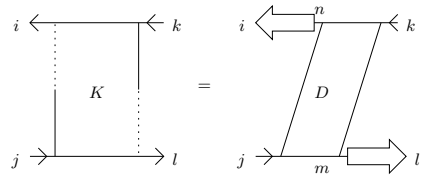}}}\\
  \subfigure[$ L $]{
    \scalebox{.7}{\includegraphics{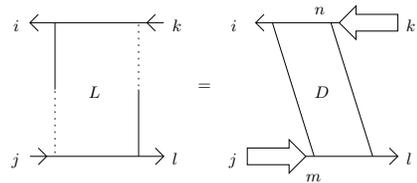}}}
  \caption{Diagrams for $ K$ and $ L$.}
  \label{fig:def-k-l}
\end{figure}

\begin{table}[htbp]
  \centering
  \begin{tabular}{c|c}
    Propagator & Explicitly bonded bases \\
\hline
    $ D_{ij;kl} $ & $ i$, $j$, $k$, and $l$ \\
    $ E_{ij;kl} $ & $ i$ and $ j$ \\
    $ K_{ij;kl} $ & $ k $ and $ j $ \\
    $ L_{ij;kl} $ & $ l $ and $ i $ 
  \end{tabular}
  \caption{Four-indexed propagators}
  \label{tab:gl-prop}
\end{table}

\pagebreak

\section{Recursion Relation and Vertices} 
\label{sec:verts}

\noindent
In \cite{oz}, two independent methods were presented that allowed one
to calculate a partition function from (\ref{eq:zder}), the integral
over $ A $. The first method was steepest descent, reviewed in
section~\ref{sec:matrix}, and which was carried to completion in
\cite{poz}. The second method, involving recursively integrating out
rows and columns of $ A $, was presented in section~VI of \cite{oz}.
\begin{align}
Z_{L1} &= Z_{L-1,1} \\  
          &+ \sum_{J=1}^{L-1} V_{LJ}     
         \expect{M_{L,J+1}^{-1}(x)}    
         \expect{M_{J,1}^{-1}(x)} \label{eq:hf-z} \\   
          &- \frac{1}{N} \sum_{J=1}^{L-1} V_{LJ}    
         \expect{M_{L1}^{-1}(x) M_{J,J+1}^{-1}(x)}_C  
         \label{eq:ga}\\ 
          &+ \sum_{J=1}^{L-1} V_{LJ}    
         \expect{M_{L,J+1}^{-1}(x) M_{J1}^{-1}(x)}_C   
         \label{eq:sd} 
\end{align}
This method gives a recursion relation (to $\ordN{-2}$) for $ Z $
which depended on objects, called ``vertices'', which were not
calculated explicitly. Here, we will use the steepest descent in order
to find a form for the vertices that can be summed directly by a
computer. It turns out that, in order to find a usable recursion
relation, both methods must be used in concert.

In order to use (\ref{eq:hf-z}) to find the $ \ordN{-2} $ partition
function, vertex operators for the insertion of bonds which cross
other bonds are required. We can find these vertices by performing a
steepest descent calculation on the \emph{recursion relation}
presented in section VI of \cite{oz}, and then doing the fluctuation
integral. This procedure will produce a modest proliferation of terms
which can be interpreted with Feynman diagrams and then summed
explicitly.
 
The first of these corrections, (\ref{eq:hf-z}), 
\begin{equation*}
  \sum_{J=1}^{L-1}
V_{LJ} \expect{M_{L,J+1}^{-1}} \expect{M_{J,1}^{-1}}
\end{equation*}
is easy to evaluate because
\begin{equation}
  \label{eq:def-zm}
  \langle M_{k+1,j}^{-1}(x)\rangle = Z_{kj}   
\end{equation}
\noindent
We need to truncate the product at $ \ordN{-2} $. The result can be
written in terms of the $ \Zb{} $ propagator introduced in
(\ref{eq:def-zb}).
\begin{multline}
  \expect{ M^{-1}_{L,J+1} } \expect{ M^{-1}_{J1} }  =  
      Z_{L-1,J+1} Z_{J-1,1} \\  
     =  G_{L-1,J+1} G_{J-1,1} 
     +  G_{L-1,J+1} \Zb{J-1,1}  \\ 
      + \Zb{L-1,J+1} G_{J-1,1} + \ordN{-4} 
  \label{eq:zz}
\end{multline}

Next we calculate (\ref{eq:ga}) by doing the fluctuation integral $
\expect{M_{L1}^{-1} M_{J,J+1}^{-1}} $ by Wick contraction, as detailed
in appendix~\ref{sec:ints}. This yields
\begin{widetext}
  \begin{equation}
   \expect{M_{L1}^{-1} M_{J,J+1}^{-1}}_{C} = 
       -\sum_{i,\dots,l} G_{L-1,i+1} G_{j-1,1} D_{ij;kl} G_{k-1,J+1}  
       G_{J-1,l+1} = -\sum_{ij} G_{L-1,i+1} \Gamma_{ij}^J G_{j-1,1} 
       \label{eq:def-gamma-old} 
\end{equation} 
\end{widetext}
Here, we have introduced the amputated vertex operator $ \Gamma $,
which can be defined very simply in terms of $ E $.
\begin{equation}
  \label{eq:def-gamma}
  \Gamma_{ij}^J = E_{ij;J+1,J-1} 
\end{equation}

\begin{figure}
  \includegraphics[width=0.20\textwidth]{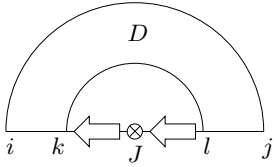}
   \caption{Diagram of $\Gamma_{ij}^J$}
   \label{fig:gam}
\end{figure}

The remaining term in the recursion relation is
\begin{widetext}
  \begin{equation}
\expect{M_{L,J+1}^{-1} M_{J1}^{-1}}_{C} =     
       \left\langle \exp \left(     
           - \sum_{p=3}^\infty \frac{(-i)^{p} T_p}    
           {p N^{p/2-1}} \right) \right. 
         \left. \left[ \left( \sum_{q=0}^\infty    
          \frac{(-i)^q B_q}{N^{q/2}}\right) G\right]_{L,J+1}   
          \nonumber \left[ \left( \sum_{r=0}^\infty  
          \frac{(-i)^r B_r}{N^{r/2}}\right) G \right]_{J1} 
          \right\rangle_{0,C} \label{eq:th-sums}
\end{equation}
\end{widetext}
This expands out to
\begin{multline}
\label{eq:th-def}
\frac{1}{N^2} \bigg\langle (B_1 G) (B_3 G) + (B_2 G) (B_2 G)    
   + (B_3 G) (B_1 G)  \\  
  +  \frac{T_3}{3} \left[(B_1 G) (B_2 G)
   + (B_2 G) (B_1 G) \right]  \\
   + \left[ \frac{T_4}{4} + \frac{T_3^2}{9} \right] 
   (B_1 G) (B_1 G) \bigg\rangle_{0,C}   
\end{multline} 
\noindent
We suppres the indices on these terms because they are all the
same: $ (B_p G)_{L,J+1} (B_q G)_{J-1,1} $. These can be evaluated the
same way that the terms in (\ref{eq:def-gamma-old}) were, by Wick
contracting the powers of $ x $ implicit in the $ T $s and $ B $s.
This is detailed in appendix~\ref{sec:ints}. Diagrams representing
each term are shown in fig.~\ref{fig:fin}, and explicit summations are
contained in appendix~\ref{sec:sums}.

These are modifications of diagrams calculated in \cite{poz}, with an
unsaturated nucleotide than can accept a hydrogen bond at $J$.  Not
all diagrams in \cite{poz} are modified this way, nor are all
insertion points included, because doing so would produce pseudoknots
of $ \ordN{-4} $.  We have further simplified notation by suppressing
the $ G $s and angle brackets. It is also convenient to discuss
``amputated'' vertices, that is, the sum of all the terms in
(\ref{eq:th-def}) and (\ref{eq:def-gamma-old}) with the external $
G_{L-1,i} $ and $ G_{j-1,1} $ propagators removed.  This amputation
insures that there are bonds attaching to the subscript indices of
both $ \Theta_{ij}^J $ and $ \Gamma_{ij}^J $.

\begin{align}
 \label{eq:rec-z}
Z_{L1}  &= Z_{L-1,1} \\  
        & +  \sum_{J=1}^{L-1} V_{LJ}     
        \left(Z_{L-1,J+1} G_{J-1,1} + G_{L-1,J+1} Z_{L-1,1}\right)
        \label{eq:rec-z-zb}\\   
        & +  \frac{1}{N} \sum_{J=1}^{L-1} \sum_{mn} 
          V_{LJ} G_{L-1,m+1} \Gamma_{mn}^J G_{n-1,1} \\ 
        & +  \sum_{J=1}^{L-1} \sum_{mn} 
           V_{LJ} G_{L-1,m+1} \Theta_{mn}^J G_{n-1,1} \label{eq:rec-th} 
\end{align}

Using the definitions for the vertex operators $ \Gamma_{ij}^J $ and $
\Theta_{ij}^J $, as well as the relation (\ref{eq:rec-z}), we can
recursively calculate the partition function for an RNA that contains
a single pseudoknot. However, we can easily extend this equation so
that it consistently accounts for RNAs that have any number of
pseudoknots. This is done by ``filling in'' the $ G $ propagators in
(\ref{eq:rec-z}) terms of the relation, to turn them into $ Z $s with
no penalty to efficiency.

\begin{figure}[htbp]
  \centering
  \subfigure[$B_1 B_3$]{
    \scalebox{0.7}{
      \includegraphics{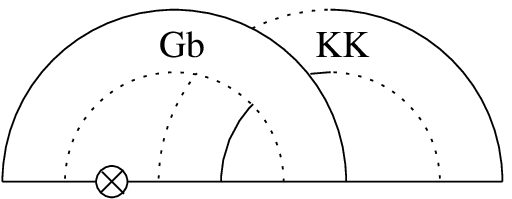}}
    \label{fig:fin:b2b2}}  
  \subfigure[$B_2 B_2$]{
    \scalebox{0.7}{
      \includegraphics{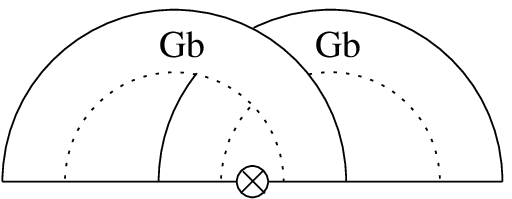}}
    \label{fig:fin:b2b2}}
  \subfigure[$B_3 B_1$]{
    \scalebox{0.7}{
      \includegraphics{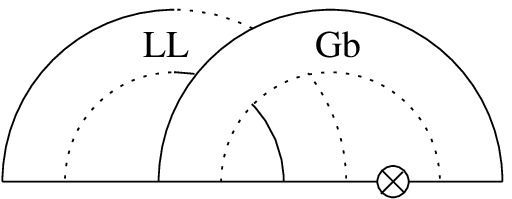}}
    \label{fig:fin:b3b1}}
  \subfigure[$B_1 B_2 T_3$]{
    \scalebox{0.7}{
      \includegraphics{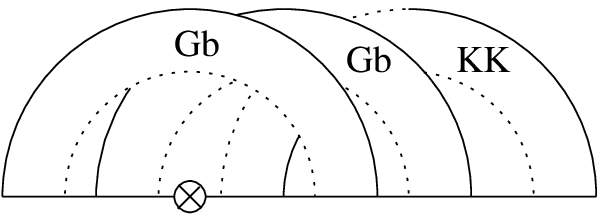}}
    \label{fig:fin:b1b2t3}}
  \subfigure[$B_2 B_1 T_3$]{
    \scalebox{0.7}{
      \includegraphics{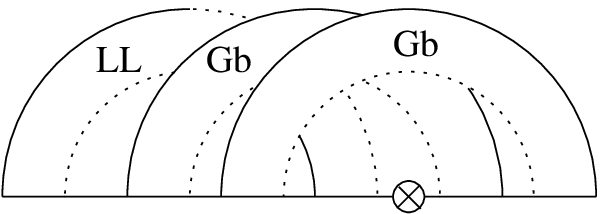}}
    \label{fig:fin:b2b1t3}}
  \subfigure[$B_1 B_1 T_4$]{
    \scalebox{0.7}{
      \includegraphics{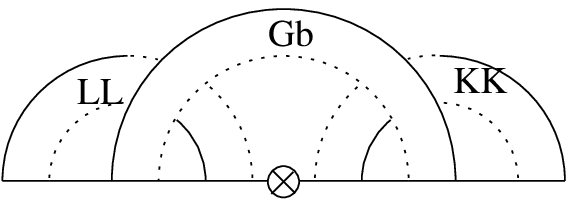}}
    \label{fig:fin:b1b1t4}}
  \subfigure[$B_1 B_1 T_3^2$]{
    \scalebox{0.7}{
      \includegraphics{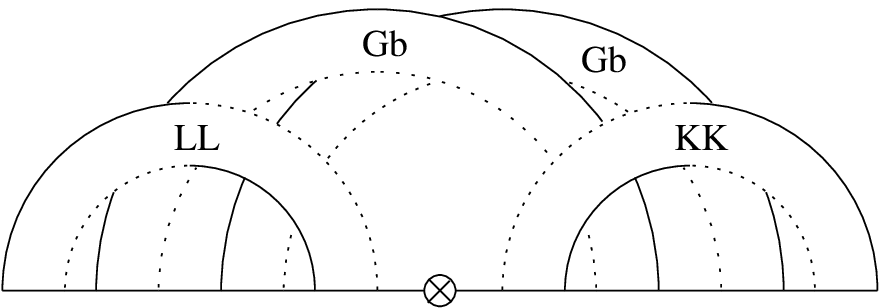}}
    \label{fig:fin:b1b1t3t3}}
  \caption{Distinct terms in $ \Theta_{ij}^J $}
  \label{fig:fin}
\end{figure}

\begin{figure}[htbp]
  \centering
  \scalebox{0.5}{\includegraphics{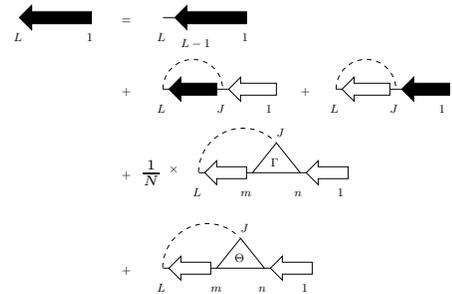}}
  \caption{Diagram form of recursion.}
  \label{fig:rec-vert-eq}
\end{figure}

\begin{align}
\label{eq:arb-pk}
Z_{L1} & = Z_{L-1,1} \\  
         & + \sum_{J=1}^{L-1} V_{LJ}     
         Z_{L-1,J+1} Z_{J-1,1}\\   
         & + \frac{1}{N} \sum_{J=1}^{L-1} \sum_{mn} 
           V_{LJ} G_{L-1,m+1} \Gamma_{mn}^J G_{n-1,1}\\ 
         & + \sum_{J=1}^{L-1} \sum_{mn} 
           V_{LJ} G_{L-1,m+1} \Theta_{mn}^J G_{n-1,1}
\end{align}

\section{Conclusion}   
\label{sec:conc}
   
\noindent   
It may be apparent that the execution time of the algorithm presented  
here grows as $ L^6 $, because the limiting step is the calculation of
the various vertices. Though a few of them actually require more than
the three indices and three summations implied by that order, those
sums can be restricted in a manner similar to that used to calculate
the four-indexed propagators. 

The model for bonding we use is somewhat simplistic, in that we
account only for independent bonds. However, it should not be
difficult to incorporate stacking interactions, or loop penalties, or
other more advanced scoring methods into the algorithm, while
retaining the topologies and recursion relation obtained from the
large $ N $ analysis.

\newpage

\appendix

\section{Explicit Vertices}
\label{sec:sums}

The $ \mathcal{O} (N^{-2}) $ vertex term, $ \Theta^J_{ij} $ is, as
shown in section~\ref{sec:verts}, a sum of several terms. While these
terms are shown in diagrammatic form in fig.~\ref{fig:fin}, we also
present the explicit summations in this appendix, to simplify efforts
to implement the algorithm.

\begin{widetext}
  
\begin{align}
  \expect{B_1 B_3}_0  = & 
     \sum_{l=j}^{J-2} \sum_{k=l+1}^{J-1} 
     E_{i,l+1;J+1,k-1} K_{J-1,j;kl} \\
  \expect{B_2 B_2}_0  = & 
     \sum_{l=j}^{J-2} \sum_{k=J+2}^{i} 
     E_{i,l+1;k,J-1} E_{k-1,j;J+1,l} \\
  \expect{B_3 B_1}_0  = &
     \sum_{l=J+1}^{i-1} \sum_{k=l+1}^{i}
     L_{i,J+1;kl} E_{k-1,j;l+1,J-1} \\
  \expect{B_1 B_2 T_3}_0  = & 
     \sum_{l=j}^{J-3} \sum_{n=l+1}^{J-2} 
     \sum_{m=n+1}^{J-1} \sum_{k=J+1}^i
     E_{i,n+1;k,m-1} E_{k-1,l+1;J+1,n} K_{J-1,j;ml} \\
  \expect{B_2 B_1 T_3}_0  = & 
     \sum_{l=j}^{J-2} \sum_{n=J+2}^{i-2} 
     \sum_{m=n+1}^{i-1} \sum_{k=m+1}^i
     L_{i,J+1,k,n-1} E_{k-1,l+1;m,J-1} E_{m-1,j;nl} \\
  \expect{B_1 B_1 T_4}_0  = & 
     \sum_{l=j}^{J-3} \sum_{n=l+1}^{J-2}
     \sum_{m=J+2}^{i-1} \sum_{k=m+1}^{i} 
     L_{i,J+1;k,m-1} E_{k-1,l+1;mn} K_{J-1,j;n+1,l} \\
  \expect{B_1 B_1 T_3^2}  = &  
     \sum_{l=j}^{J-4} \sum_{n=l+1}^{J-3} \sum_{p=n+1}^{J-2}
     \sum_{o=J+2}^{i-2} \sum_{m=o+1}^{i-1} \sum_{k=m+1}^{i}
     L_{i,J+1;k,o-1} E_{k-1,n+1;mp} \\
     \nonumber
      & \qquad \qquad \times E_{m-1,l+1;on} K_{J-1,j;p+1,l} 
\end{align}
\end{widetext}

\section{Evaluation of Integrals}
\label{sec:ints}

We calculate (\ref{eq:ga}) by evaluating the integral 
\begin{widetext}
  \begin{align}
   \expect{M_{L1}^{-1} M_{J,J+1}^{-1}}_C & = 
      \Bigg\langle \exp \left( 
        -\sum_{p=3}^\infty \frac{(-i)^p}{pN^{p/2-1}}T^p
      \right) \left[ \left( \sum_{q=0}^\infty 
          \frac{(-i)^q}{N^{q/2}} B_q\right)G\right]_{L1} 
      \left[ \left(\sum_{r=0}^\infty   
          \frac{(-i)^r}{N^{r/2}} B_r\right)G\right]_{J,J+1} 
      \Bigg\rangle_{0,C} \\ 
      & =  \Bigg\langle\left( 
        1-\frac{i\,T_3}{3N^{\half}}   
        - \frac{T_4}{4N} - \frac{ T_3^2}{18N} \right) \left[
      \left(1- \frac{i\,B_1}{N^{1/2}}    
          - \frac{B_2}{N} \right) G \right]_{L1} 
      \left[ \left(  
          1- \frac{i\,B_1}{N^{\half}} -\frac{B_2}{N} \right)  
        G \right]_{J,J+1} \Bigg\rangle_{0,C} 
\end{align}
\end{widetext}
We have only expanded to $ \ordN{-1} $ because of the explicit factor of
$ 1/N $ in (\ref{eq:ga}). The $\ordN{-\half} $ terms vanish by
symmetry, since they depend on odd powers of $ x $, so this simplifies
to
\begin{widetext}
  \begin{multline}
  \label{eq:ga-terms}
  \expect{M_{L1}^{-1} M_{J,J+1}^{-1}}_C  = 
    \Bigg\langle G_{L1}G_{J,J+1}\left[ 
      1-\frac{1}{N}\left(
        \frac{T_4}{4}+\frac{T_3^2}{18} \right) \right]
      - \frac{T_3}{3 N} \left[ 
        (B_1 G)_{L1} G_{J,J+1} + G_{L1} (B_1 G)_{J,J+1} \right]\\
      - \frac{1}{N} \left[ 
        (B_2 G)_{L1} G_{J,J+1} + G_{L1} (B_2 G)_{J,J+1} \right]
       - \frac{1}{N} (B_1 G)_{L1} (B_1 G)_{J,J+1} \Bigg\rangle_{0,C} 
\end{multline}
\end{widetext}

\begin{figure}[htbp]
  \centering 
  \subfigure[$\langle T_3 B_1 \rangle$]{ \scalebox{0.5}{
      \includegraphics{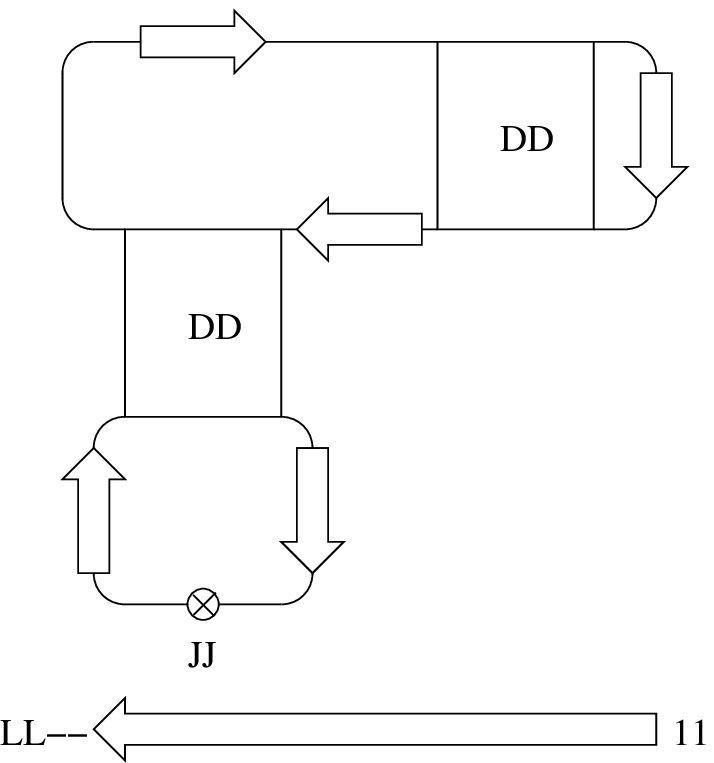}
      \label{fig:vangam:a}}}
  \subfigure[$\langle B_2 \rangle$]{ \scalebox{0.5}{
      \includegraphics{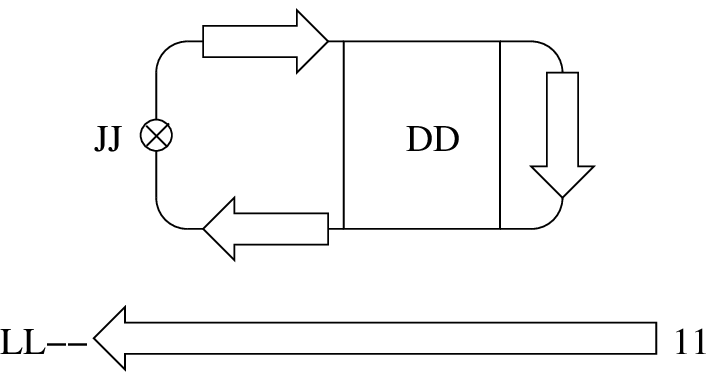}
      \label{fig:vangam:b}}}
   \caption{Vanishing contractions in $\Gamma$}
   \label{fig:folds}
\end{figure}

The $\ord{1}$ term $ \expect{G_{L-1,1} G_{J-1,J+1}} $ vanishes because
it depends on the ``backwards propagator'' $ G_{J-1,J+1} $. Four
$\ordN{-1}$ terms, $\expect{T_4 G_{L-1,1} G_{J-1,J+1}}$, $
\expect{T_3^2 G_{L-1,1} G_{J-1,J+1}} $, $ \expect{T_3 (B_1 G)_{L1}
  G_{J-1,J+1}} $ and $ \expect{(B_2 G)_{L1} G_{J-1,J+1}} $, vanish for
the same reason.  This leaves us with only three terms to compute, $
\expect{T_3 G_{L-1,1} (B_1 G)_{J,J+1}}$, $ \expect{G_{L-1,1} (B_2
  G)_{J,J+1}}$ and $ \expect{(B_1 G)_{L1} (B_1 G)_{J,J+1}}$.  The
first two are shown in diagram form in figs.  \ref{fig:vangam:a} and
\ref{fig:vangam:b} respectively; they obviously vanish. The third is
shown as a diagram in fig.  \ref{fig:gam} and can be written as an
explicit sum,
\begin{widetext}
  \begin{align}
   \expect{(B_1 G)_{L1} (B_1 G)_{J,J+1}}_0  =  
       \sum_{i,\dots,l} G_{L-1,i+1} G_{j-1,1} D_{ij;kl} G_{k-1,J+1} 
       G_{J-1,l+1} =  \sum_{ij} G_{L-1,i+1} E_{ij;J+1,J-1} G_{j-1,1}
\end{align}
\end{widetext}
The left hand side depends on $ L $ while the right hand side depends
on $ L-1 $ because of the non-standard definition of matrix products
that we introduced in section~\ref{sec:matrix}.

The same integration technique applies to (\ref{eq:th-sums}).
\begin{widetext}
\begin{multline} 
    \expect{M_{L,J+1}^{-1} M_{J1}^{-1}}_C =  \left\langle \left(1 -
    \frac{i\,T_3}{3N^{1/2}} - \frac{T_4}{4N} - 
    \frac{T_3^2}{18N} \right) \right. \\
    \times \left[ \left(1-
    \frac{i\,B_1}{N^{1/2}} - \frac{B_2}{N} + 
    \frac{i\,B_3}{N^{3/2}} \right) G \right]_{L, J+1}       
    \left. \left[ \left(1 - \frac{i\,B_1}{N^{1/2}} - \frac{B_2}{N} +
    \frac{i\,B_3}{N^{3/2}} \right) G \right]_{J1}   
    \right\rangle_{0,C} 
\end{multline}
\end{widetext}
We drop some terms in the expansion because they cannot yield
non-vanishing connected expectation values.  The remaining terms can
be expanded to $\ordN{-2}$. The fractional coefficients cancel
over-counting in terms that depend on traces, which are symmetric
under cyclic permutations of the $ x $s.

As an example, we evaluate the first of the terms by Wick contraction.
The contractions which remain are
\begin{widetext}
  \begin{align}
    \nonumber
  \expect{ (B_2 G)_{L,J+1}(B_2 G)_{J,1} }_{0,C} &= 
   \sum_{i,\dots,p}   
   G_{L-1,i+1} G_{k-1,m+1} G_{o-1,J+1} G_{J-1,p+1}G_{n-1,l+1}
   G_{j-1,1} (D_{in;kp}D_{mj;ol} +D_{io;km}D_{pj;nl})     
  \\ & =   
    \sum_{i,\dots,l} G_{L,i+1} G_{j-1,1} E_{il;k+1,J-1}E_{kj;J+1,l-1} 
  \end{align}
\end{widetext}

\begin{figure}[hbtp]
   \centering
   \subfigure[Non-vanishing contraction]{ 
      \scalebox{0.6}{ 
      \includegraphics{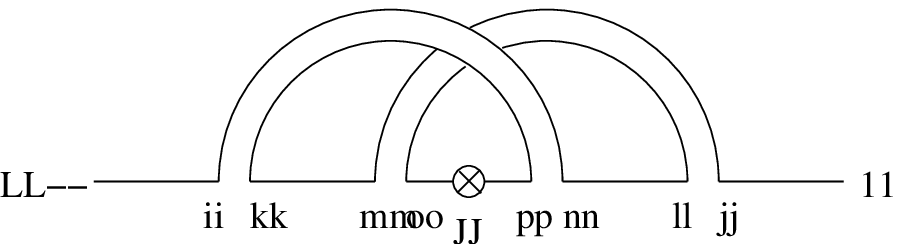} 
      \label{fig:thb2b2:a}}} 
   \subfigure[Vanishing contraction.]{ 
      \scalebox{0.6}{ 
      \includegraphics{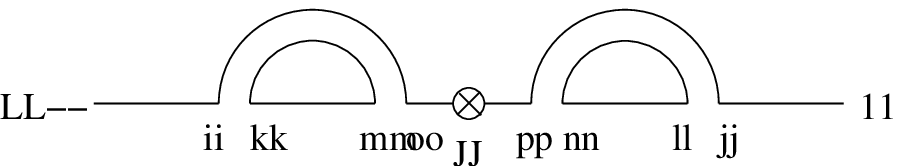}
      \label{fig:thb2b2:b}}} 
   \caption{Contractions in $\langle B_2 B_2 \rangle$}
   \label{fig:thb2:b} 
\end{figure}

The first of the two Wick contractions is shown as a diagram in fig.
\ref{fig:thb2b2:a}, while the second term, shown in fig.
\ref{fig:thb2b2:b}, vanishes. The non-vanishing part can be greatly
simplified with the use of the appropriate propagators. The advantage
of the propagators introduced in section \ref{sec:props} should now be
obvious; we have reduced the number of summed indices from eight to
four.

\begin{thebibliography}{99}

\bibitem{oz} H. Orland and A. Zee, ``RNA Folding and Large N Matrix
    Theory'', \verb+cond-mat/0106359+.
\bibitem{poz} M. Pillsbury, H. Orland and A. Zee, ``A steepest
    descent calculation of RNA'', \verb+physics/0207110+.
\bibitem{nj} R. Nussinov and A.B. Jacobson, PNAS \textbf{77} (1980) 6309.
\bibitem{coleman} S. Coleman, \textit{Aspects of Symmetry} 
    (Cambridge University Press, New York, 1985), Ch. 9.
\bibitem{higgs} P.G. Higgs, Quarterly Reviews in Biophysics
    \textbf{33} (2000) 199.
\bibitem{tinoco} I. Tinoco Jr. and C. Bustamante,
    J. Mol. Biol. \textbf{293} (1999) 271.
\bibitem{smith} M.S. Waterman and T.F. Smith, Adv. Applied Maths. 
    \textbf{7} (1986)
\bibitem{rivas} E. Rivas and S.R. Eddy, J. Mol. Biol. \textbf{285}
    (1999) 2053.    
\bibitem{thooft} G. \mbox{'t Hooft}, Nucl. Phys. B \textbf{72} (1974)
    461.
\bibitem{zuker} M. Zuker, Science \textbf{244} (1989) 48.
\bibitem{jaeger} J.A. Jaeger, D.H. Turner and M. Zuker,
  Proc. Nat. Acad. Sci., \textbf{86} (1989) 7706. 
\bibitem{ln} D.K. Lubensky and D.R. Nelson, 
    Phys. Rev. Lett. \textbf{85} (2000) 1572. 
\bibitem{hofetal} I.L. Hofacker, W. Fontana, P.F. Stadler, 
    L.S. Bonhoeffer, M. Tacker and P. Schuster, Monatshefte f\"ur
    Chemie  
    \textbf{125} (1994) 167. 
\bibitem{mm} A. Montanari and M. M\'ezard,
  Phys. Rev. Lett. \textbf{86} 
    (2001) 2178.
\bibitem{bh} R. Bundschuh and T. Hwa, Phys. Rev. Lett. \textbf{83} 
    (1999) 1479. 
\bibitem{mccaskill} J.S. McCaskill, Biopolymers 29 (1990) 1105. 
\bibitem{zzo} H. Zhou, Y. Zhang and Z-C. Ou-Yang, Phys. Rev. Lett. 
    \textbf{86} (2001) 356. 

\end{thebibliography}
\end{document}